\documentclass[intlimits,twoside,a4paper]{article}

\usepackage{amsmath,amssymb}
\usepackage{graphicx}
\usepackage{float}

\usepackage[T2A]{fontenc}
\usepackage[cp1251]{inputenc}

\usepackage[eqsecnum]{cmpj2}
\issue{2014}{17}{1}{13802}
\doinumber{10.5488/CMP.17.13802}

\title[The effect of surface scattering]%
{The effect of electron surface scattering on fine metal particle electromagnetic radiation absorption}
\author[I.A. Kuznetsova, M.E. Lebedev, A.A. Yushkanov]{I.A. Kuznetsova\refaddr{label1}, M.E. Lebedev\refaddr{label1}, A.A. Yushkanov\refaddr{label2}}
\addresses{
\addr{label1} Department of Microelectronics, Yaroslavl State University, 14 Sovetskaya St., 150003  Yaroslavl, Russia
\addr{label2} Department of Theoretical Physics, Moscow State Regional University,  10a Radio St., 105005 Moscow, Russia
}

\date{Received July 24, 2013, in final form November 5, 2013}
\authorcopyright{I.A. Kuznetsova, M.E. Lebedev, A.A. Yushkanov, 2014}

\begin{document}

\maketitle

\begin{abstract}
The magnetic dipole absorption cross section of a spherically shaped metal particle was calculated in terms of kinetic approach. The particle considered was placed in the field of a plane electromagnetic wave. The model of boundary conditions  was investigated taking into account the dependence of the reflectivity coefficient both on the surface roughness parameter and on the electron incidence angle. The results obtained were compared with theoretical computation results for a model of combined diffusion-specular boundary conditions of Fuchs.
\pacs 78.67.Bf, 42.25.Bs
\end{abstract}

\section{Introduction}

The well-known fact is that optical and electromagnetic properties of fine particles (size is much smaller than the wavelength of electromagnetic field) could significantly differ from the properties of bulk samples \cite{c1,c2,c3,c4,c5}. To explain these size effects it is usually necessary to use the quantum approach \cite{c4,c5}, but in some cases one can also use the kinetic method  \cite{c1,c2,c3}. When the spherical particle radius $ a$ is compared, or smaller than the mean free path of electrons $ \lambda$, the interactions between conduction electrons and the surface beneficially effect the particle optical characteristics. Therefore, such optical quantities as the absorption cross section become non-trivial functions of the ($ a \leqslant \lambda$) ratio. In such cases, classical theory (Mie theory \cite{c6}) based on the equations of macroscopic electrodynamics is not applicable but the problem can be solved in terms of the kinetic approach.

Nowadays, one can produce particles several nanometers in size. The mean free path of electrons in ordinary metals with high conductivity (Au, Ag, Cu, Al etc.) is about $10\div100$ nanometers, and the de~Broil is compatible with the atomic spacing ($ \Lambda_\textrm{B} \approx 0.3$~nm) \cite{c7}. If the condition  $ \Lambda_\textrm{B} \sim a < \lambda$ is satisfied, the classical kinetic description of size effects is applicable  which is determined by quasi-classical motion of electrons \cite{c7}.
Composite materials containing metal nanoparticles attract the attention of researchers.
Thin films are used for military purposes as screen covers. Besides, they increase the efficiency of solar cells \cite{c8}. The absorption capability of such materials greatly depends on the size, shape and origin of the particles embedded. It is necessary to study the mechanism of electromagnetic radiation absorption by a single nanoparticle in order to describe the composite optical properties. There are a lot of unsolved problems in studying this mechanism, such as experimental discovery of anomalous high absorption by metal particles in far-infrared region (several orders higher than theoretical predictions) \cite{c9,c10}. Several theoretical methods were suggested to explain the anomalous behavior of absorption cross section.

The present paper is focused on the effect of electron scattering mechanism on electromagnetic radiation absorption cross section in a fine spherical metal particle.

In the problems of this type, the contribution of the surface electron scattering is based on step-by-step formulation of a boundary condition to the Boltzmann equation which relates the distribution functions of the incident electrons and the electrons reflected by the surface. In fact, the boundary condition replaces the collision integral. In the kinetic problem of thin film resistance, Fuchs formulated the boundary condition \cite{c11} having introduced the reflectivity coefficient  $ q$ which means the relative amount of electrons specularly reflected by the surface. Coefficient $ q$ can also be represented as the probability of specular reflection of the electrons ($0<q<1$) and $ (1-q)$ means the probability of diffusion reflection, respectively.
In \cite{c2}, the authors proposed a step-by-step kinetic description of the size effect on the infra-red magnetic dipole absorption in case of diffusion reflecting  $ (q=0)$ in a small spherical metal particle. A number of papers contain an assumption which associates the anomalous high infra-red absorption with a possible prevalence of specular electron reflection. However, in \cite{c12}  it is shown that taking into account a mixed diffusion-specular boundary condition with a variable coefficient  $ q$ it is possible just to reduce (by one order) but not to annihilate the incompatibility between theoretical and experimental results (by two-three orders) \cite{c13,c14}.

Several experimental and theoretical papers are focused on the relationship between the reflectivity coefficient and surface properties  \cite{c15,c16,c17}.

The model of boundary conditions considering the coefficient $ q$ as a function of the incidence angle  $ \gamma$ between the electron velocity vector and the particle radius was investigated in \cite{c15}. For quasi-sliding electrons, when the angle $ \gamma$ is close to $ \pi/2$, the quantity $ 1-q$ is proportional to  $ \cos^2{\gamma} $ [$(q-1) \sim \cos^2{\gamma}$]~\cite{c15}.

The model is useful in many kinetic applications. For example, it was used in calculating the static electric conductivity of a thin round wire \cite{c18}.

In the present paper, there was chosen a Soffer model of boundary conditions which considers the dependence of $ q$ both on the surface imperfection and on the incidence angle $ \gamma$. Non-equilibrium distribution function describing the conduction electron response to the alternating magnetic field of a plane electromagnetic wave is calculated using the kinetic approach for a metal sphere having the radius  $ a $. The ratio $ a/ \lambda$ was not limited. The results obtained are compared with theoretical calculations for Fuchs model of boundary conditions  \cite{c2,c12}.

\section{Formulation of the problem}

Small metal spherically shaped particle is placed into the plane electromagnetic field with frequency $ \omega$. The range of possible frequencies can be obtained by the condition $ \omega \ll \omega_\textrm{p}$, where $ \omega_\textrm{p}$ is plasma resonance frequency. In other words, the maximum frequency is limited by the near infrared range. To the point, the problem of optical properties of nanoparticles near the plasmon frequency was discussed in numerous works. For example, a detailed kinetic description of the behavior of surface plasmons in spherical metal particles one can see in \cite{c19}.

The radius $ a$ is smaller than the skin-depth $ \delta $ and consequently the skin-effect is negligible. As noted above, the ratio $ a/ \lambda$ is not limited. In \cite{c2}, the authors provide detailed estimates which show that for $1\div10$~nm particles, the contribution of the dipole electric polarization current is negligible compared to the contribution of vortex currents induced by the magnetic field of the wave due to the screening of the electric field in the particle in this range of frequencies. Therefore, the effect of an external electric field is not considered.

The process of electromagnetic radiation absorption can be represented as the generation of the electric field $ \mathbf{E}$ by the alternating uniform magnetic field  $ \mathbf{H}=\mathbf{H_0} \exp{(-\ri \omega t)}$. From Maxwell’s equations one has:
\begin{equation}
\label{eq:1}
\mathbf{E}=\frac{1}{2c}\left [ \mathbf{r} \frac{\partial \mathbf{H}}{\partial t}\right ] = \frac{\omega}{2 \ri c} \left [ \mathbf{r} \mathbf{H}_0 \right ] \exp{(-\ri \omega t)},
\end{equation}
where $ \omega$ stands for the angular frequency of the wave, $ c$ is the speed of light,  $ \mathbf{H}_0 $ is the magnetic field amplitude, $ \mathbf{r}$ is the electron position vector (origins in the centre of the particle).

The electric field $ \mathbf{E}$ causes the vortex current  $ \mathbf{j}$. The current can be obtained using the local Ohm’s law, when $ a \gg \lambda$:
\begin{equation}
\label{eq:2}
\mathbf{j}=\Sigma (\omega) \mathbf{E},\qquad \Sigma (\omega)=\Sigma_0 /(1-\ri \tau \omega),
\end{equation}
where $ \Sigma (\omega)$ stands for Drude conductivity, $ \Sigma_0 = e^2 n \tau /m $ is static conductivity,  $ e$ is electron charge, $ n$ is density of electrons, $ m$ is mass of electrons, $ \tau$ is relaxation time.

However, when the particle radius $ a $ is compatible with $ \lambda$, the relationship between $ \mathbf{j}$ and $ \mathbf{E}$ shows a non-local behavior, and thus macroscopic equations cannot be implemented.

The electric field in equation ( \ref{eq:1}) causes electron distribution function to deviate from the equilibrium Fermi function $ f_0$:
\begin{equation}
\nonumber
f( \mathbf{r,v,t}) = f_0( \varepsilon )+ f_1( \mathbf{r,v,t}),
\end{equation}
where $ \mathbf{v}$ stands for the electron velocity. Then, one can consider the electron kinetic energy via a classical formula  $ \varepsilon = mv^2/2$, and use a step-like approximation for equilibrium distribution function  $ f_0 (\varepsilon)$:
\begin{equation}
\label{eq:3}
f_0( \varepsilon)= \theta ( \varepsilon_\textrm{F} - \varepsilon)= \left\{
\begin{array}{ll}
 1 , \quad 0 \leqslant \varepsilon \leqslant \varepsilon_\textrm{F}\,, \\
 0 , \quad \varepsilon > \varepsilon_\textrm{F}\,,
\end{array} \right.
\end{equation}
where  $ \varepsilon_\textrm{F} = m v_\textrm{F}^2/2$ is Fermi energy ($ v_\textrm{F}$ is Fermi velocity). Fermi surface is supposed to have spherical shape.

The current density in the particle can be represented as:
\begin{equation}
\label{eq:4}
\mathbf{j} = e \int{\mathbf{v} f \cfrac{2 \rd^3(m v)}{h^3}}=2 e {\cfrac{m}{h^3}}^3 \int{\mathbf{v} f_1 \rd^3 v}.
\end{equation}

The mean power per second dissipated in the particle can be found as \cite{c20}:
\begin{equation}
\label{eq:5}
\overline{Q}=\int{\overline{ \Re( \mathbf{E})\cdot\Re( \mathbf{j}) } \, \rd^3r}=\frac{1}{2}\Re\int{ \mathbf{j}\cdot \mathbf{E^\ast} \, \rd^3r},
\end{equation}
where overlining means the time averaging, asterisk means complex conjugation. The absorption cross section can be obtained by dividing the mean power per second, $ \sigma$ is equal to the average power $ \overline{Q}$ in equation (\ref{eq:5}) being dissipated by the mean energy flux of the incident wave:
\begin{equation}
\label{eq:6}
\sigma=\frac{\overline{Q}}{(cE_0^2/8\pi)}\,.
\end{equation}

\section{Absorption cross section}

Function $ f_1$ satisfies the kinetic equation in the linear approximation according to the field:
\begin{equation}
\label{eq:7}
-\ri \omega f_1+ v \frac{\partial f_1}{\partial r}+e(\mathbf{v\cdot E})\frac{\partial f_0}{\partial \varepsilon}=-\frac{f_1}{\tau}\, ,
\end{equation}
where $ f_1 \propto \exp (-\ri \omega t) $ and collision integral is calculated during the relaxation time approximation ($ \tau$ stands for the average period of collisions).

One can find a unique solution of equation (\ref{eq:7}) if the boundary condition for non-equilibrium function $ f_1(\mathbf{r,v},t) $ is formulated. There was used a condition taking into account the dependence of the reflectivity coefficient $ q$ both on the surface imperfection and on the angle $ \gamma $ between the electron velocity and the particle radius \cite{c15}:
\begin{equation}
\label{eq:8}
f_1(\mathbf{r,v})=q(H, \cos{\gamma}) \, f_1(\mathbf{r,v'}), \qquad \text{ when}\quad  |\mathbf{r}|=a, \quad  \mathbf{r \cdot v}<0,
\end{equation}
\begin{equation}
\label{eq:9}
q(H, \cos{\gamma})=\exp{ \left[ -(4 \pi H)^2 \cos^2{\gamma} \right] }, \qquad H=h_s/ \lambda_\textrm{F}\,,
\end{equation}
where $ \mathbf{v'}=\mathbf{v}-2\mathbf{r} (\mathbf{r \cdot v})/a^2 $ is the velocity vector prior to the collision, $ \mathbf{v} $ is the velocity vector after the collision; $ h_s $  is the mean square relief height; $ \lambda_\textrm{F} $ means de~Broil electron wavelength on the Fermi surface.
Kinetic equation  (\ref{eq:7}) can be solved in terms of characteristic method \cite{c21}. The increment of position vector along the characteristic (trajectory) could be written as follows:
\begin{equation}
\label{eq:10}
 \rd \mathbf{r}= \mathbf{v}\rd t,
\end{equation}
the change of the function $ f_1 $ is given by
\begin{equation}
\label{eq:11}
\rd f_1=-\left[ \nu f_1+e(\mathbf{v\cdot E})\frac{\partial f_0}{\partial \varepsilon}\right] \rd t' ,
\end{equation}
where
\begin{equation}
\label{eq:12}
\nu =\frac{1}{\tau}-\ri \omega
\end{equation}
is complex scattering frequency.

The boundary condition in equation (\ref{eq:8}) allows one to follow the change of the nonequlibrium distribution function $ f_1$ along the trajectory being reflected. Function $ f_1$ has a discontinuity in the point of reflection:
\begin{equation}
\label{eq:13}
f_1(t'_n+0)=q(H, \cos{\gamma})f_1(t'_n-0),
\end{equation}
where  $ n$ is reflection index, $ t'_n$ is parameter value of $ n$-th collision (see figure~1).

Angular momentum is conserved for a specular reflection. Thus, along the trajectory:
\begin{equation}
\label{eq:14}
[\mathbf{rv}]=\textrm{const},
\end{equation}
and the difference $ (t'_n-t'_{n-1}) $ is independent of $ n$:
\begin{equation}
\label{eq:15}
t'_n=nT+\textrm{const},
\end{equation}
where $ T$ stands for time of flight between the points $ r_{n-1}$ and $ r_n$:
\begin{equation}
\label{eq:16}
T=-2(\mathbf{r_n \cdot v_n})/v^2.
\end{equation}

The expression $ \mathbf{(E \cdot v)}$ is also constant:
\begin{equation}
\label{eq:17}
\mathbf{E \cdot v} = \frac{\omega}{2 \ri c} \left[ \mathbf{rH} \right] \cdot \mathbf{v}=\frac{\omega}{2 \ri c} \left[ \mathbf{rv} \right] \cdot \mathbf{H}=\textrm{const}.
\end{equation}
The relationship between the function $ f_1$ values at two adjacent points can be found from equation (\ref{eq:11}) using the conditions from equation (\ref{eq:13}):
\begin{equation}
\label{eq:18}
f_1(t'_n+0)=q \left\{ -\frac{e\mathbf{(E \cdot v)}}{\nu} \, \frac{\partial f_0}{\partial \varepsilon} \left[ 1-\exp(- \nu T) \right] +f_1(t'_{n-1}+0) \exp(- \nu T)   \right\}.
\end{equation}
One can express $ f_1(t'_{n-1}+0) $ via  $ f_1(t'_{n-2}+0) $ and repeat the iterations. As a result, one obtains $ f_1(t'_n+0) $ as a sum of geometric progression with the denominator $ q  \exp(- \nu T) $:
\begin{equation}
\label{eq:19}
 f_1(t'_n+0)=-q \cfrac{-\cfrac{e\mathbf{(E \cdot v)}}{\nu} \, \cfrac{\partial f_0}{\partial \varepsilon}  \left[ 1-\exp(- \nu T) \right] }{1- q \exp(- \nu T)} \,.
\end{equation}
After integrating the equation (\ref{eq:11}) with the initial condition from equation (\ref{eq:19}) one has:
\begin{equation}
\label{eq:20}
 f_1(t')=\cfrac{e\mathbf{(E \cdot v)}}{\nu} \, \cfrac{\partial f_0}{\partial \varepsilon} \left[  \cfrac{(1-q)\exp(- \nu t')}{1-q \exp(- \nu T)} -1 \right].
\end{equation}
Parameters $ t' $ and $ T$ can be related with the point coordinates $ \mathbf{(r,v)} $ in the phase space by  \cite{c12}:
\begin{equation}
\label{eq:21}
\mathbf{r}=\mathbf{r_0}+\mathbf{v}t', \qquad \mathbf{r_0 \cdot v}<0, \qquad \mathbf{r_0}^2=a^2, \qquad T=-2(\mathbf{r_0 \cdot v})/v^2.
\end{equation}
Here, parameter $ t'$ means the time of electron motion from the collision point to $ \mathbf{r}$ with the speed $ \mathbf{v}$. One can exclude $ \mathbf{r_0}$ from equations (\ref{eq:21}):
\begin{equation}
\label{eq:22}
t'=\left[ (\mathbf{r \cdot v})+ \sqrt{(\mathbf{r \cdot v})^2+(a^2-r^2)v^2} \right] \Big/v^2, \qquad T=2 \sqrt{(\mathbf{r \cdot v})^2+(a^2-r^2)v^2}\Big/v^2.
\end{equation}

\begin{figure}[htb]
\centerline{
\includegraphics[width=0.3\textwidth]{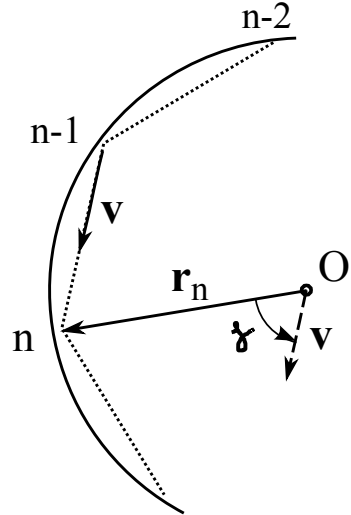}
}
\caption{Specular reflection of the electron from the inner surface of the particle, $ \mathbf{r}_n $~---~radius-vector of $ n$-th reflection point.} \label{fig:1}
\end{figure}

Equations (\ref{eq:20})--(\ref{eq:22}) describe the function $ f_1(\mathbf{r,v}) $ completely. The obtained function allows one to calculate the current in equation (\ref{eq:4}).
To calculate the current it is convenient to use spherical coordinates both for coordinate space  ($r,\, \theta, \, \phi $, polar axis $ z \, || \, \mathbf{H_0}$) and for velocity space  ($ v, \beta, \alpha$, the polar axis is $ \mathbf{ v_r}$). The electric field $ \mathbf{E} $ has only $ \phi $-component in spherical coordinates:
\begin{equation}
\label{eq:23}
\mathbf{E}=\mathbf{e_{\phi}}E_\phi\,, \qquad
E_\phi=\frac{\ri \omega}{2c}rH_0 \sin\theta \exp(-\ri \omega t)\,.
\end{equation}
The current  $ \mathbf{j}$ also has only $ \phi $-component (vortex current lines are perpendicular to $ z$):
\begin{equation}
\label{eq:24}
j_\phi = \cfrac{2\pi e^2 m^2 E_\phi v_\textrm{F}^3 }{h^3 \nu} \left\{ \frac4 3 + \int_0^\pi \cfrac{[q(H, \cos \gamma)-1] \exp(-\nu t')}{1 - q(H, \cos \gamma) \exp(-\nu T) } \sin ^3 \alpha \, \rd \alpha \right\}.
\end{equation}
It is useful to make the substitution:
\begin{equation}
x_0 = \cfrac{a}{v_\textrm{F} \tau}\,, \qquad  y_0 = \cfrac{a \omega}{v_\textrm{F}}\, , \qquad z_0 = \cfrac{a}{v_\textrm{F}} \nu = \cfrac{a}{v_\textrm{F}} \left(\cfrac 1 \tau - \ri \omega\right) = x_0 - \ri y_0\,, \qquad \xi = \cfrac r a\,, \qquad \mu =  \cfrac {\mathbf{v \cdot r}}{v r} = \cos \alpha\,,
\nonumber
\end{equation}
\begin{equation}
\nonumber
v t' = z_0 \cfrac{v_\textrm{F} t'}{a}  = z_0 \left( \xi \mu + \sqrt{\xi^2 \mu^2 +1 - \xi^2 } \right) = z_0 \left( \xi \mu + \eta_0 / 2 \right),
\end{equation}
\begin{equation}
\nonumber
 vT = z_0 \cfrac{ v_\textrm{F} T}{a}  = 2 z_0  \sqrt{\xi^2 \mu^2 +1 - \xi^2 }= z_0 \eta_0\, , \qquad \cos{\gamma} = \cfrac{vT}{2a} = \sqrt{\xi^2 \mu^2 +1 - \xi^2 } = \cfrac{\eta_0 }{2}\,.
\end{equation}
Dimensionless variables $ x_0$, $y_0$, $z_0 $ are normalized to the Fermi velocity $ v_\textrm{F}$.
After transformations the current is as follows:
\begin{equation}
\label{eq:25}
j_\phi = \cfrac{2\pi e^2 m^2 E_\phi v_\textrm{F}^3 }{h^3 a} \cfrac{1}{z_0} \left( \cfrac4 3 + \int_{-1}^1 \cfrac{\left\{ \exp \left[- (2 \pi H \eta_0)^2 \right] -1 \right\} \exp \left[ -z_0 \left( \xi \mu +{ \eta_0}/{2}\right) \right] }{1 - \exp \left[- (2 \pi H \eta_0)^2 \right] \exp(-z_0 \eta_0) } (1-\mu^2) \, \rd\mu \right).
\end{equation}
Absorption cross section in equation (\ref{eq:6}) can be found by substituting equation (\ref{eq:25}) into equation (\ref{eq:6}):
\begin{equation}
\label{eq:26}
\sigma=\sigma_0F(H,x_0,y_0),
\end{equation}
where
\begin{equation}
\nonumber
\sigma_0= \cfrac{\pi^2 n e^2 a^4 v_\textrm{F}}{2 m c^3}\,, \qquad n=2 \cfrac {4\pi}{3} \left( \cfrac{m}{h} \right)^3 \cfrac{\mathbf{v}_\textrm{F}^3}{a^3}\,,
\end{equation}
\begin{eqnarray*}
 F(H,x_0,y_0) &=&\Re \left[ \cfrac{4y^2_0}{z_0} \left( \cfrac{4}{15} + \int_0^1\int_{-1}^{1}  \cfrac{\left\{ \exp \left[- (2 \pi H \eta_0)^2 \right] -1 \right\} \exp \left[ -z_0 \left( \xi \mu +  {\eta_0}/{2}\right) \right] }{1 - \exp \left[- (2 \pi H \eta_0)^2 \right] \exp(-z_0 \eta_0) }
 \right.\right.\\
&&{}\times\left.\left.
 \xi^4 (1-\mu^2)  \, \rd\xi \, \rd\mu \vphantom{\int_{0}^1}\right) \right].
\end{eqnarray*}

The integral can be transformed to the single one by introducing $ \rho = \sqrt{ \xi^2 \mu^2 +1 - \xi^2 }$, $u=\xi \mu $  and the expression reduces to a single integral:
\begin{equation}
\label{eq:27}
F(H,x_0,y_0) = \Re \left\{ \cfrac{4y^2_0}{z_0} \left[ \cfrac{4}{15}+\int_0^1 \rho (1-\rho^2) \cfrac { \left[ \exp \left(- (4 \pi H \rho)^2 \right) -1 \right] \left[ 1 - \exp ( -2z_0 \rho)\right] }{z_0 \left[1-\exp \left(- (4 \pi H \rho)^2 \right) \exp ( -2z_0 \rho) \right]} \, \rd \rho \right] \right\}.
\end{equation}

The dimensionless absorption cross section is the function of the electron inverse mean free path $ x_0 $, the dimensionless field frequency $ y_0 $ and the roughness parameter $ H $.

\begin{figure}[!b]
\centerline{
\includegraphics[width=0.5\textwidth]{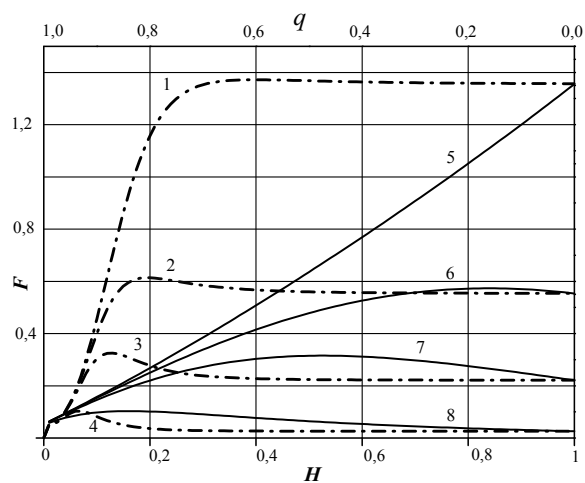}
}
\caption{Dimensionless absorption cross section $ F $ vs roughness parameter $ H $ for the model boundary conditions of Soffer (solid lines 1, 2, 3, 4 --- $ y_0 = 2, 1, 0.6, 0.2 $, respectively) and the dependence $ F(q) $ for the model of Fuchs (dashed lines: 5, 6, 7, 8 --- $ y_0 = 2, 1, 0.6, 0.2 $, respectively) at $ x_0 = 0.1 $.} \label{fig:2}
\end{figure}

\section{Limiting cases}

Macroscopic asymptote can be implemented at $ x_0 \gg 1 $ for any reflecting mechanism [in this case, in equation (\ref{eq:27}), one can neglect exponential members due to their rapid decay]
\begin{equation}
\label{eq:28}
 F_\textrm{as} =\cfrac{16y_0^2}{15} \Re \left( \cfrac{1}{x_0-  \ri y_0} \right)= \cfrac{16 x_0 y_0^2}{15(x_0^2+y_0^2)}\,.
\end{equation}

The equation remains valid in case of the surface being suitably prepared, when $ h_s $ tends to zero. In this case, the mechanism of interactions is completely specular, and coefficient  ($q \to 1$).

This fact could be related to the phenomenon that the boundary does not effect the distribution function $f $. High-frequency current satisfies the local Ohm’s law along the specular-reflecting trajectory with any ($ a \leqslant \lambda$) ratio.

\begin{figure}[!t]
\centerline{
\includegraphics[width=0.5\textwidth]{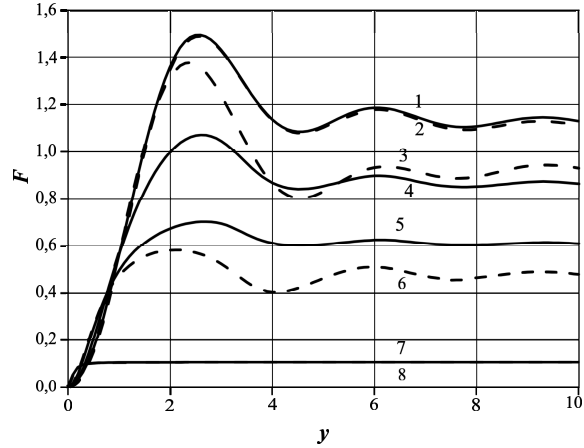}
}
\caption{Dimensionless absorption cross section $ F$ vs dimensionless frequency $ y_0 $. Solid lines for the boundary conditions of Fuchs 1, 4, 5, 7 --- $q = 0, 0.25, 0.5, 1 $, respectively. Dashed lines for the boundary condition of Soffer 2, 3, 6, 8 --- $ H = 1, 0.25, 0.1, 0 $, respectively. $ x_0 = 0.1 $ for each curve.} \label{fig:3}
\end{figure}

\section{Results and discussion}

\begin{figure}[!b]
\centerline{
\includegraphics[width=0.5\textwidth]{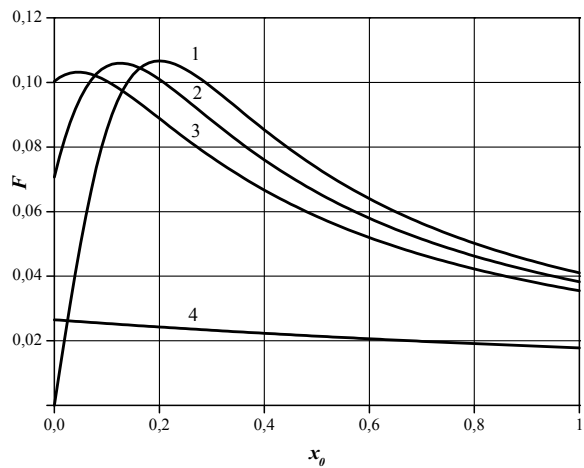}
}
\caption{Dimensionless absorption cross section $ F$ vs dimensionless inverse mean free path  $ x_0 $.  The model of boundary conditions of Soffer: curves 1,2,3,4 --- $ H = 0, 0.04, 0.06, 1 $, respectively. $ y_0 = 0.2 $ for each curve. } \label{fig:4}
\end{figure}

\begin{figure}[!t]
\centerline{
\includegraphics[width=0.5\textwidth]{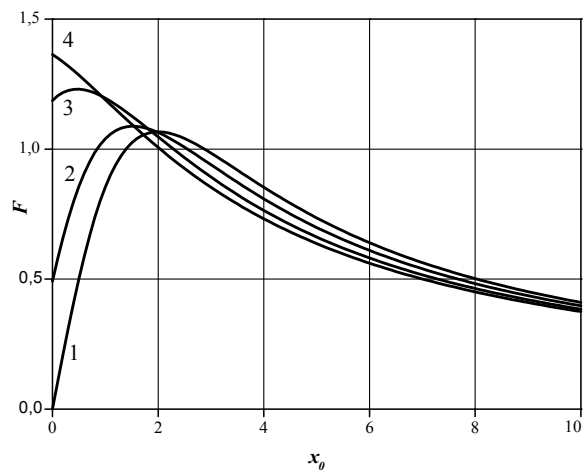}
}
\caption{Dimensionless absorption cross section $ F$ vs dimensionless inverse mean free path  $ x_0 $. The model of boundary conditions of Soffer: curves 1,2,3,4 --- $ H = 0, 0.04, 0.06, 1 $, respectively. $ y_0 = 2 $ for each curve. } \label{fig:5}
\end{figure}

The comparison of dimensionless absorption cross-sections $ F$ calculated for the model of boundary conditions of Soffer and for the model of boundary conditions of Fuchs is illustrated in figure~\ref{fig:2}. Dimensionless inverse mean free path $ x_0 $ equals 0.1 for all curves, dimensionless field frequency  $ y_0 $ changes from 0.2 to 2.
Spectra both for the model of boundary conditions of Soffer and for the model of boundary conditions of Fuchs are shown in figure~\ref{fig:3}. Dimensionless inverse mean free path  $ x_0 $ equals 0.1 for all curves. It is obvious that the Soffer model provides a more oscillating spectrum behavior.
The dependencies of absorption cross-section on the dimensionless inverse mean free path $ x_0 $ are drawn in figures~\ref{fig:4} and~\ref{fig:5}. Dimensionless field frequency $ y_0 = 2 $ equals 0.2 for curves in figure \ref{fig:4} and 2 for all curves in figure~\ref{fig:5}. The function has a maximum for certain $ x_0 $  (it depends on $ H $). The contrast is the highest for curves 1 ($ H = 0 $) and is achieved when the dimensionless field frequency $ y_0 $ is equal to the dimensionless frequency of surface collisions $ x_0 $. The maximum shifts to lower values of $ x_0 $ with an increase of parameter $ H $. Vertical shift of the maximum depends from the value $ y_0 $. For low frequencies ($ y_0 < 0.9$, figure~\ref{fig:4}), the maximum goes down when parameter $ H $ increases. For high frequencies  ($ y_0 > 1$, figure~\ref{fig:5}), the maximum goes up when parameter $ H $ increases.  One can also see that all curves tend to a macroscopic asymptote in equation (\ref{eq:28}) for high values of $ x_0 $.

\section{Conclusions}

The considered classical size effects  act beneficially on the magnetic dipole absorption cross-section. The origin of these effects is associated with the mechanism of surface scattering of electrons for particles with the mean free size of electrons or less.
The probability of specular reflection increases for quasi-sliding electrons. Using the model of boundary conditions of Soffer it was shown how this phenomenon acted on the size effect behavior.
Magnetic dipole absorption cross-section significantly depends on the surface properties, in particular, on the finish quality. These facts can be explained by the relationship of reflectivity coefficient with roughness parameter. Depending on the field frequency (for relative high frequencies:  $ y_0> 1$), the maximum of absorption decreases with an increase of the particle size and an increase of roughness parameter.
The results obtained are in good agreement with theoretical results obtained for the model of diffusion-specular boundary conditions of Fuchs \cite{c12} in limiting cases of specular reflecting surface and diffusion reflecting surface.

\section*{Acknowledgement}
This work was performed at the center for facilities sharing ``Micro-
and nanostructures diagnostics'' supported by the Ministry of Education
and Science.

\ukrainianpart

\title{Вплив розсіяння електронів поверхнею на електрмагнітну випромінювальну адсорбцію дисперсної металічної частинки}
\author{І.А. Кузнєцова,\refaddr{label1}, М,Є. Лєбєдєв\refaddr{label1}, A.A. Юшканов\refaddr{label2}}

\addresses{
\addr{label1} Кафедра мікроелектроніки, Ярославський державний університет,  150003  Ярославль, Росія
\addr{label2} Кафедра теоретичної фізики, Московський  державний обласний університет, 105005 Москва, Росія
}

\makeukrtitle

\begin{abstract}
\tolerance=3000%
Переріз магнітної дипольної адсорбції металічної частинки сферичної форми розраховано в термінах кінетичного підходу.
Частинка, що розглядається, була поміщена в поле плоскої електромагнітної хвилі.
Досліджувалась модель граничних умов, що враховує залежність коефіцієнта відбивання як від параметра жорсткості поверхні, так і від
кута налітання електрона. Отримані результати порівняно з результатами теоретичних розрахунків для моделі комбінованих
дифузійно-дзеркальних граничних умов Фучса.

\keywords переріз електромагнітної адсорбції, рівняння Больцмана,
граничні умови Соффера, резонансноподібне явище
\end{abstract}

\end{document}